\documentclass[a4paper,twocolumn,nofootinbib,nobibnotes,aps,prl,floatfix,superscriptaddress,10pt,longbibliography,reprint]{revtex4-2}
\usepackage[utf8]{inputenc}
\usepackage[english]{babel}
\usepackage{amsmath}
\usepackage{amsthm}
\usepackage{amssymb}
\usepackage{graphicx}
\usepackage[caption=false]{subfig}
\usepackage{xparse}
\usepackage{xcolor}
\usepackage[separate-uncertainty=true, exponent-product=\cdot]{siunitx}
\usepackage{hyperref}
\usepackage{braket}

\definecolor{forestgreen}{rgb}{0.13, 0.55, 0.13}

\begin{document}

\title{Generalised quantum interference}

\author{Sebastian Malewicz}
\affiliation{Australian Research Council Centre of Excellence for Engineered Quantum Systems}
\affiliation{School of Mathematics and Physics, University of Queensland, Brisbane, Australia}

\author{Laura Serino}
\affiliation{School of Mathematics and Physics, University of Queensland, Brisbane, Australia}

\author{Marcelo P. de Almeida}
\affiliation{Australian Research Council Centre of Excellence for Engineered Quantum Systems}
\affiliation{School of Mathematics and Physics, University of Queensland, Brisbane, Australia}

\author{Markus Rambach}
\affiliation{Australian Research Council Centre of Excellence for Engineered Quantum Systems}
\affiliation{School of Mathematics and Physics, University of Queensland, Brisbane, Australia}

\author{Ming~Su}
\affiliation{Australian Research Council Centre of Excellence for Engineered Quantum Systems}
\affiliation{School of Mathematics and Physics, University of Queensland, Brisbane, Australia}
\affiliation{Silux Technologies Co., Ltd., Shanghai 200233, China}

\author{Jihun Cha}
\affiliation{Australian Research Council Centre of Excellence for Engineered Quantum Systems}
\affiliation{School of Mathematics and Physics, University of Queensland, Brisbane, Australia}
\affiliation{QICI Quantum Information and Computation Initiative, School of Computing and Data Science, The University of Hong Kong, Pok Fu Lam Road, Hong Kong}

\author{Till J. Weinhold}
\affiliation{Australian Research Council Centre of Excellence for Engineered Quantum Systems}
\affiliation{School of Mathematics and Physics, University of Queensland, Brisbane, Australia}

\author{Austin Lund}
\affiliation{School of Mathematics and Physics, University of Queensland, Brisbane, Australia}
\affiliation{Australian Research Council Centre of Excellence for Quantum Computation and Communications Technology}
\affiliation{Dahlem Center for Complex Quantum Systems, Freie Universit\"at Berlin, Berlin, Germany}

\author{A.~G. White}
\affiliation{Australian Research Council Centre of Excellence for Engineered Quantum Systems}
\affiliation{School of Mathematics and Physics, University of Queensland, Brisbane, Australia}

\begin{abstract}
Hong-Ou-Mandel interference is one of the clearest signatures of quantum behaviour: two identical photons meeting at a beam splitter always leave together. Here we demonstrate that this symmetry can be broken.
Using a programmable integrated photonic processor, we interfere imbalanced two-photon states on a variable beam splitter and show that the output state of quantum interference can be continuously tuned, reaching the extreme case where destructive interference suppresses bunching at one output port.
The asymmetry creates a forbidden bunching channel: photons may exit together at one port or separately at both ports, but cannot emerge together at the other port.
\end{abstract}

\maketitle

The Hong-Ou-Mandel (HOM) effect \cite{hong1987, ou1987, fearn1987, rarity1989} is a quintessential manifestation of quantum interference \cite{bouchard2020, drago2024}. When two indistinguishable photons impinge on different input ports of a balanced beam splitter, they interfere such that they always leave together. Since this behaviour depends directly on photon indistinguishability, HOM interference has evolved from a foundational demonstration of nonclassical physics into a standard tool for assessing indistinguishability \cite{slussarenko2017, raymer2010, kobayashi2016, karimi2014,hiekkamaki2021a}, characterising photon sources \cite{kiraz2005, kaltenbaek2006, beugnon2006, maunz2007, mosley2008a, sanaka2009, bernien2012, senellart2017, Morrison2022} and optical components and interfaces \cite{politi2008, zhu2022a, babel2023, gera2024}, and performing Bell-state analysis \cite{kwiat1998, calsamiglia2001, bayerbach2023}. Beyond characterisation, HOM-based techniques have also found practical application in high-resolution imaging and microscopy \cite{ndagano2022, torre2023}, precision metrology \cite{lyons2018, chen2019, meskine2024c} and quantum-enhanced sensing \cite{nasr2003, basiri-esfahani2015, lyons2023}.

Nevertheless, the familiar HOM effect is not the only possibility for two-photon quantum interference. In the standard case, destructive interference suppresses the coincidence outcome, in which the photons leave in different output ports. More generally, however, linear optics allows the interference to be controlled by changing the beam splitter transformation, enabling generalised quantum interference \cite{lund2015}. In this framework, destructive interference can be tuned to the extreme case in which the suppressed output is instead one of the two bunching terms, giving rise to asymmetric quantum interference. For suitably prepared two-photon input states, the photons can emerge separately or bunch in one output port, but not bunch in the other.

Here we experimentally demonstrate generalised quantum interference on a programmable integrated photonic processor \cite{maring2024}. We prepare imbalanced two-photon states and interfere them on a variable beam splitter, showing how the full two-photon output distribution can be reshaped by tuning the input-state imbalance and the beam splitter reflectivity. Within this family, we realise the asymmetric quantum interference regime in which one bunching output is suppressed while the remaining bunching output and the coincidence output are equally probable.

\begin{figure}
    \centering
    \includegraphics{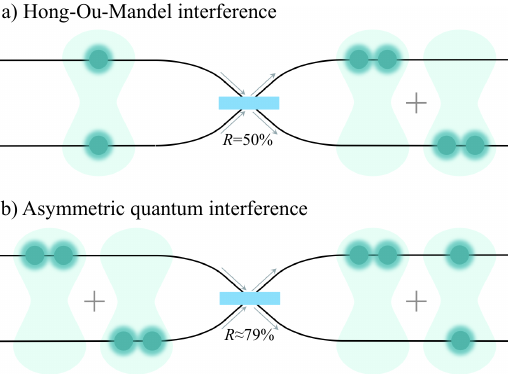}
    \caption{Conceptual comparison of Hong-Ou-Mandel (HOM) interference and generalised two-photon quantum interference at the asymmetric point. 
    (a) In HOM interference, two indistinguishable photons entering different input ports of a balanced beam splitter interfere such that the coincidence output is suppressed, leaving a superposition of the two bunched outputs.
    (b) In asymmetric quantum interference, an imbalanced two-photon input interferes on a beam splitter such that one of the two bunching outputs is suppressed, while the coincidence output and the other bunching output remain equally probable.}
    \label{fig:concept}
\end{figure}

For two photons in two modes---such as the two possible output paths of a beam splitter---the possible output configurations are $\ket{1,1}$, in which the photons leave in different ports, and $\ket{2,0}$ and $\ket{0,2}$, in which they bunch into the same port. In the standard HOM effect, two indistinguishable photons entering different input ports of a balanced beam splitter interfere such that the $\ket{1,1}$ term is suppressed, leaving a superposition of the two bunched outputs (Fig.~\ref{fig:concept}a). HOM interference is therefore associated with destructive interference in the coincidence channel.

Ref.~\cite{lund2015} showed that the standard HOM effect is only one particular case of a broader class of generalised two-photon interference effects accessible through linear optics. More generally, two-photon states can be classified into equivalence classes, such that two states in the same class can be transformed into each other using only linear optical elements. Each class has a canonical form, which in the two-mode case can be written as the superposition of bunched states
\begin{equation}
    \label{eq:canonical}
    a\ket{2,0}+\sqrt{1-a^2}\ket{0,2} \,,
\end{equation}
with $0\leq a\leq 1$. By preparing a two-photon input state in this form and then interfering it on a variable beam splitter, as shown in Fig.~\ref{fig:concept}, one can access any target state within the same equivalence class. This means that, starting from input states with different imbalance coefficients $a$ and varying the beam splitter reflectivity and the phase shifts, one can continuously tune the relative weights and phases of the three output terms, thereby obtaining an arbitrary two-photon, two-mode state \cite{lund2015}.

In particular, for $a=(\sqrt{3}+1)/2\sqrt{2}\approx0.966$ and beam splitter reflectivity $R=1/(3-\sqrt{3})\approx78.9\%$ with a transmission phase shift of $\pi/2$, the output becomes
\begin{equation}
    \ket{\psi_{\mathrm{out}}}=\frac{1}{\sqrt{2}}\left(\ket{2,0}+i\ket{1,1}\right),
\end{equation}
so that the $\ket{0,2}$ term is suppressed, while the $\ket{2,0}$ and $\ket{1,1}$ terms are equally probable (Fig.~\ref{fig:concept}b). By symmetry, the complementary reflectivity $R\approx21.1\%$ would suppress the opposite bunching term $\ket{2,0}$.

\begin{figure}
    \centering
    \includegraphics[width=\linewidth]{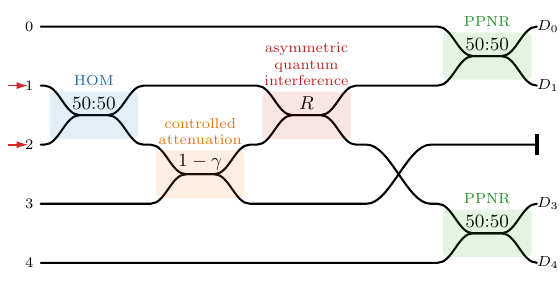}
    \caption{Optical circuit implemented in the photonic processor Ascella. Two single photons interfere on a 50:50 beam splitter to produce a two-photon bunched state. A beam splitter with reflectivity $1-\gamma$ introduces loss in one of the two channels, thereby preparing the unbalanced input state of Eq.~\eqref{eq:canonical}, which is interfered on a variable beam splitter with reflectivity $R$. The three possible two-photon outputs are resolved using pseudo-photon-number-resolving (PPNR) detectors implemented with a 50:50 beam splitter in each output channel and detector pairs $D_0$-$D_1$ and $D_3$-$D_4$, following the mode numbering of the implemented subcircuit.}
    \label{fig:circuit}
\end{figure}

To realise generalised quantum interference experimentally, we use the programmable photonic quantum processing unit (QPU) Ascella \cite{maring2024}. We program the chip such that two single photons are first interfered on a balanced beam splitter to generate a HOM bunched state, after which controlled attenuation in one arm prepares the imbalanced input of Eq.~\eqref{eq:canonical}. This state is then interfered on a variable beam splitter, and the three two-photon outputs are resolved using pseudo-photon-number-resolving (PPNR) detection.

The five-mode subcircuit of Ascella used for this experiment is shown in Fig.~\ref{fig:circuit}. We begin with two single photons generated by Ascella's quantum dot source and temporal demultiplexer \cite{maring2024}. Interfering these photons on a balanced beam splitter produces the typical HOM bunched state, and controlled attenuation of one arm then prepares the imbalanced superposition. After postselection on two-photon events, the resulting state at this point takes the form of Eq.~\eqref{eq:canonical} with $a=\left[1+(1-\gamma)^2\right]^{-1/2}$, where $\gamma$ is the transmissivity of the beam splitter that couples one arm to a loss channel. The state is then sent to a variable beam splitter with reflectivity $R$. 

To resolve the three possible two-photon outputs, we use PPNR detection at the outputs of the variable beam splitter. Each output arm is split on an additional balanced beam splitter and monitored by a pair of superconducting nanowire detectors, so that coincidences within one arm identify bunched events, while clicks in different arms identify the $\ket{1,1}$ outcome. Since this scheme detects bunched two-photon events with half the efficiency of coincidence events, we correct for this known imbalance when extracting the output probabilities.

\begin{figure}
    \centering
    \includegraphics{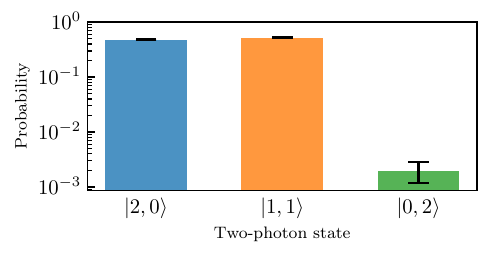}
    \caption{Output-state probabilities at the asymmetric interference point ($\gamma=0.7327$ and $R\approx78.9\%$). Bars (shown on a logarithmic scale) represent the theoretical probabilities, while markers indicate the measured values; error bars denote $1\sigma$ statistical uncertainties. The results show a $~250$-fold suppression of the $\ket{0,2}$ term with respect to $\ket{2,0}$ and $\ket{1,1}$.
    }
    \label{fig:barplot}
\end{figure}

Fig.~\ref{fig:barplot} shows the measured output probabilities at the asymmetric interference point, corresponding to $\gamma=0.7327$ and $R\approx0.789$. Under these conditions, one bunching output is strongly suppressed, with measured probability $P(\ket{0,2})=\num{1.9\pm0.8e-3}$, while the remaining two outputs have similar probabilities, $P(\ket{2,0})=\num{.479\pm0.009}$ and $P(\ket{1,1})=\num{.519\pm0.009}$. The suppressed $\ket{0,2}$ output therefore occurs with a probability approximately 250 times smaller than the other two outputs. This is the signature of asymmetric quantum interference: destructive interference suppresses one bunching output of the two-photon output state, leaving the other two outputs equally probable.

\begin{figure}
    \centering
    \includegraphics{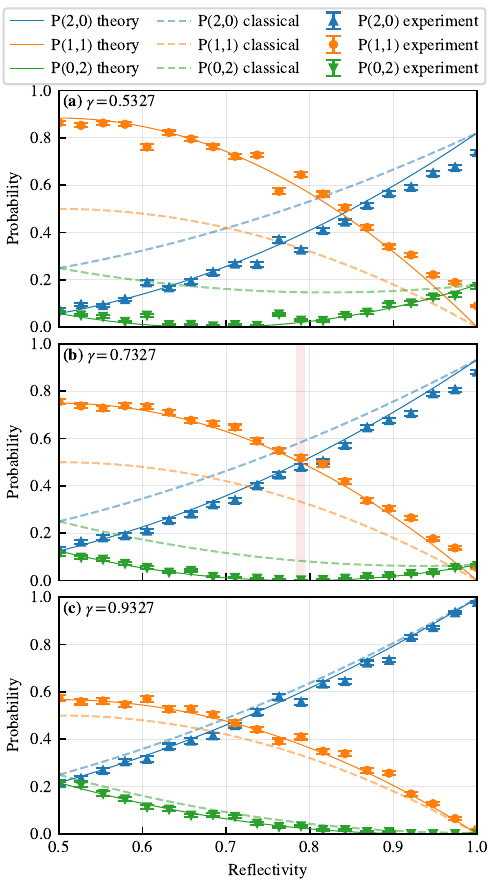}
    \caption{Output-state probabilities as a function of the variable beam splitter reflectivity $R$ for three input-state imbalance settings: (a) $\gamma=0.5327$, (b) $\gamma=0.7327$, and (c) $\gamma=0.9327$. Markers show the measured probabilities for the three two-photon output states $\ket{2,0}$, $\ket{1,1}$, and $\ket{0,2}$, with error bars denoting $1\sigma$ statistical uncertainties; solid lines show the corresponding theoretical predictions, and dashed lines show the predictions for pairs of classical particles, calculated as the decohered version of Eq.~\eqref{eq:canonical} \cite{lund2015}. The shaded region in Fig.~(b) indicates the conditions for asymmetric quantum interference, where $P(0,2)=0$ and $P(2,0)=P(1,1)$, shown in detail in Fig.~\ref{fig:barplot}.}
    \label{fig:results}
\end{figure}

To observe the broader generalised interference behaviour and show how the system evolves into and out of the asymmetric interference point, we program three attenuation settings, $\gamma=0.5327$, $0.7327$, and $0.9327$, and for each of them scanning 20 beam-splitter reflectivities between $R=0.5$ and $R=1$. Fig.~\ref{fig:results} shows the measured probabilities for the $\ket{2,0}$, $\ket{1,1}$, and $\ket{0,2}$ outputs, together with the corresponding theoretical predictions. For comparison, we also plot the predictions obtained by treating the unbalanced two-photon input state in Eq.~\eqref{eq:canonical} as completely decohered, corresponding to the limit of poor mode overlap where the inputs can be regarded as pairs of classical particles \cite{lund2015}.

As $R$ is varied, the output probabilities evolve continuously, in close agreement with the theoretical predictions. For each value of $\gamma$, there is a reflectivity $R$ at which the $\ket{0,2}$ term is suppressed, but in general this does not coincide with the reflectivity at which the remaining two outputs are equally probable. These two conditions coincide only for the appropriate input state imbalance, yielding the asymmetric interference point.
The measurements therefore show the generalised interference behaviour directly: varying the input-state imbalance and beam splitter reflectivity redistributes probability among all three two-photon output configurations, with the asymmetric point appearing as the extreme case.
The deviation from the classical particle prediction highlights the genuinely quantum origin of this interference.

The small discrepancies between the measured and predicted probabilities can be attributed to imperfections in the implemented transformation, including finite photon indistinguishability and small calibration errors in the programmed circuit. These effects become more visible for reflectivities closer to unity, where the realised beam splitter is most sensitive to experimental imperfections. Higher-photon-number contributions and background noise can also be post-selected as two-photon events and slightly modify the measured statistics. Despite these limitations, the measured distributions reproduce the expected generalised-interference behaviour across the full range of programmed settings.

In conclusion, we have experimentally demonstrated generalised quantum interference on a programmable integrated photonic processor. By preparing imbalanced two-photon states and interfering them on a variable beam splitter, we show that the output distribution can be continuously reshaped and that the destructively suppressed component can be shifted away from the coincidence channel of the standard HOM effect. As a limiting case, we realise the asymmetric point, where one bunching output is suppressed by destructive interference while the remaining bunching and coincidence outputs are equally probable. The measured output distributions agree closely with the expected evolution as the input-state imbalance and beam splitter reflectivity are varied, confirming controlled access to generalised quantum interference effects. The ability to suppress a selected output channel in a controlled way may also prove useful for preparing entangled photon number states with linear optics \cite{wang2005, liu2008}. These results place HOM interference within a broader family of two-photon quantum interference effects, experimentally accessible with linear optics.

\paragraph{Acknowledgements}
We acknowledge the traditional owners of the land on which the University of Queensland is situated, the Turrbal and Jagera people.
This research was funded by the Australian Government through the Australian Research Council Centre of Excellence for Engineered Quantum Systems (EQUS, CE170100009).
AGW is the recipient of an Australian Research Council Laureate Fellowship (FL210100045) funded by the Australian Government.

\appendix

\section{Method}
The generalised quantum interference experiment is implemented on Quandela's programmable photonic QPU Ascella \cite{maring2024}. This platform combines a quantum-dot single-photon source, a temporal demultiplexer, a reconfigurable integrated interferometer, and superconducting nanowire single-photon detectors (SNSPDs).
The interferometer itself is based on a 12-mode QuiX universal photonic chip \cite{taballione2021}, although only the five-mode subcircuit with depth 5 shown in Fig.~\ref{fig:circuit} is used here.
For the measurements reported here, Ascella had an overall single-photon system efficiency (from source to measurement) of $\sim2.5\%$, a multi-photon-event suppression of $97.8\%$, and a photon indistinguishability characterised by a HOM visibility of $\sim82.8\%$.

To generate the unbalanced two-photon state, we first produce two single photons using Ascella's quantum dot and temporal demultiplexer. We inject one photon into each input 1 and 2 in Fig.~\ref{fig:circuit}, the two input ports of a balanced beam splitter. This first beam splitter generates the standard HOM output state $(\ket{2,0}+\ket{0,2})/\sqrt{2}$, where the relative phase is absorbed into subsequent phase-shifters in the circuit. We then attenuate the $\ket{0,2}$ component with a variable beam splitter coupled to a loss channel with transmissivity $\gamma$. We note that, while this operation is not unitary on the two-mode signal subspace and therefore introduces lower-photon-number terms, these contributions are removed by post-selecting only two-photon detection events. After renormalisation, the prepared two-photon input state takes the canonical form of Eq.~\eqref{eq:canonical}, with $a=(1+(1-\gamma)^2)^{-\frac{1}{2}}$.

Next, we interfere this state on a variable beam splitter with reflectivity $R$. To resolve the three possible two-photon output states, we apply pseudo-photon-number-resolving (PPNR) detection at the outputs of the variable beam splitter. We implement this by splitting each output mode on an additional balanced beam splitter and detecting the resulting four ports with four SNSPDs, two in each arm. In this arrangement, a coincidence between the two SNSPDs in the same output arm identifies a bunched event, namely $\ket{2,0}$ or $\ket{0,2}$, whereas a click in each arm indicates the coincidence outcome $\ket{1,1}$.
In the ideal case, the PPNR scheme detects $\ket{1,1}$ events with unit efficiency, but detects bunched two-photon events with probability $1/2$. We therefore correct for this known difference in detection efficiency when normalising the measured probabilities.

We program three attenuation settings in the first stage, corresponding to $\gamma= 0.5327, 0.7327, 0.9327$. For each attenuation setting, we program 20 values of the variable beam splitter reflectivity, linearly spaced between $R=0.5$ and $R=1$. 
We note that the circuit in Fig.~\ref{fig:circuit} is not implemented by controlling each element in the QPU. Instead, for each set of parameters, a clear-box optimisation algorithm is used to find the optimal circuit configuration to achieve the desired unitary transformation \cite{fyrillas2024}.

For each programmed setting, we collect $5\cdot10^5$ counts in total across the four detectors, corresponding to an acquisition time of approximately \SI{4}{\s} at a repetition rate of \SI{80}{\MHz}. We group the resulting detection events into the three outcomes $\ket{1,1}$, $\ket{2,0}$, and $\ket{0,2}$ according to the PPNR detection pattern described above, and we then normalise the counts to obtain the corresponding output probabilities. 

\bibliography{bibliography}

\end{document}